\documentclass[]{interact}

\usepackage{epstopdf}

\usepackage[numbers,sort&compress,merge]{natbib}
\bibpunct[, ]{[}{]}{,}{n}{,}{,}

\newcommand{\be}{\begin{equation}}
\newcommand{\ee}{\end{equation}}

\newcommand{\prt}{\partial}

\newcommand{\bt}{\beta}

\newcommand{\ep}{\varepsilon}
\newcommand{\al}{\alpha}
\newcommand{\ra}{\rightarrow}
\newcommand{\sgm}{\sigma}

\newcommand{\Gm}{\Gamma}

\newcommand{\Lbd}{\Lambda}

\articletype{ARTICLE TEMPLATE}

\begin{document}

\title{Self-similar summation of virial expansions}

\author{
\name{Vyacheslav ~I. Yukalov$^{a,b}$\thanks{CONTACT V.~I. Yukalov. Email: yukalov@theor.jinr.ru} 
and 
Elizaveta ~P. Yukalova$^c$}
\affil{$^a$Bogolubov Laboratory of Theoretical Physics,
Joint Institute for Nuclear Research, Dubna 141980, Russia;
$^b$Instituto de Fisica de S\~{a}o Carlos, 
Universidade de S\~{a}o Paulo, CP 369, S\~{a}o Carlos 13560-970, S\~{a}o Paulo, Brazil;
$^c$Laboratory of Information Technologies, 
Joint Institute for Nuclear Research, Dubna 141980, Russia }
}

\maketitle

\begin{abstract}
Virial expansions are the series in powers of density assumed to be small. However, the 
equations of state require to consider finite densities for which virial expansions, as 
a rule, diverge. In order to extrapolate a virial expansion to the values, where this 
expansion diverges, one uses summation methods. The most often used method is the Pad\'{e} 
summation, which has several deficiencies. First of all, Pad\'{e} approximants are not 
uniquely defined, suggesting a large table of admissible variants. Second, often there 
appear spurious unphysical poles. On the contrary, in those cases where the existence of 
a pole is physically motivated, Pad\'{e} approximants do not necessarily exhibit it. A 
new approach for the summation of virial expansions is suggested, based on self-similar 
approximation theory. The method is regular and uniquely defined. It allows for the 
determination of physically motivated poles. The accuracy of self-similar approximants is 
not worse than that of the best Pad\'{e} approximants with fitting parameters or of Monte 
Carlo simulations. The self-similar summation is based solely on virial expansions, involving
no fitting parameters. In some cases, self-similar summation allows for reconstructing the 
sought functions exactly. The approach is illustrated by summing virial expansions for 
hard-disk fluids, hard-sphere fluids, and systems with power-law potentials. 
\end{abstract}


\section{Introduction}

One of the most powerful, systematic methods for studying the properties of matter is via 
a virial expansion yielding power series with respect to density \cite{Mayer_1,Hill_2,
Morita_3,Stell_4,Mason_5,Hansen_6,Masters_7}. The radius of convergence for virial expansions 
is usually rather small, because of which the virial series as such are valid only for very 
small density \cite{Lebowitz_8,Procacci_9}, while for the equations of state in the whole 
range of the admissible density virial expansions need to be extrapolated. 

In order to extrapolate a virial expansion to the region of finite and relatively large 
densities, one resorts, as a rule, to the phenomenological constructions or to Pad\'{e} 
approximants complemented by additional fitting to experimental values. Sometimes, as
in the case of hard-sphere fluids, it is required to separate a term with a fixed pole 
imitating the existence of closest packing \cite{Carnahan_13,Gutierrez_14,Song_15,Mulero_16,
Solana_17,Hu_18,Tian_19,Tian_20,Maestre_21,Boublik_22}. 

The use of Pad\'{e} approximants has the following weak points. First, this way does not 
constitute a uniquely defined method, since for each truncated series of order $k$ there is 
a large table of $k$ admissible Pad\'{e} approximants. Second, Pad\'{e} approximants often
exhibit spurious unphysical poles. On the contrary, in order to imitate the existence of 
closest packing, the approximants have to posses a pole of the type $(1- A x)^{-\alpha}$, 
where $x$ is the packing fraction. However, not each Pad\'{e} approximant enjoys such a pole. 
In addition, when extrapolating a virial expansion, it would be desirable to have some 
theoretical justification for the chosen approximating form. 

In the present paper, we suggest a new approach for extrapolating virial expansions, which
is based on self-similar approximation theory advanced in Refs. \cite{Yukalov_23,Yukalov_24,
Yukalov_25,Yukalov_26} and recently reviewed in \cite{Yukalov_27,Yukalov_28}. In order to
introduce the approach in a clear way, we start with hard-core fluids for which the related
virial expansions are well known and the method can be straightforwardly illustrated. In 
Section II, we formulate the problem for the compressibility factor of $d$-dimensional hard 
spheres. For hard-sphere potentials, the virial coefficients do not depend on temperature, 
which allows for the development of explicit equations of state. The hard-sphere systems 
not merely are of interest as such, but they often serve as references for designing 
perturbation theory for realistic systems \cite{Hansen_6,Mulero_12}. This is because the 
properties of gases and liquids are dominated by repulsive forces associated with the 
hard-cores of particles.

The basics of the approach are explained in Section III. We sketch the main ideas, but omit 
mathematical details that can be found in the cited papers and reviews 
\cite{Yukalov_27,Yukalov_28}. We concentrate on the presentation of practical steps for 
implementing the general methodology. The advantage of the method of self-similar extrapolation 
is in the following. First, the self-similar approximants are uniquely defined. Second, they 
automatically prescribe the point of singularity, if it exists, and the related critical 
exponent. Third, the prediction of higher-order virial coefficients is also done automatically.
In addition, the derivation of the self-similar approximants has a justification in the 
detection of the property of similarity between subsequent truncated series. The form of the 
self-similar approximants, by construction, is more general then that of Pad\'{e} approximants, 
which explains why self-similar approximants, as a rule, allow for better accuracy than 
Pad\'{e} approximants, or at least not worse than the latter. 
    
An explicit application is shown in Section IV for the simple example of hard rods. We 
illustrate specifications by following the method of self-similar summation and show that 
the self-similar extrapolation of the virial expansion for hard-rod fluids gives the exact 
expression for the compressibility factor coinciding with the result of geometric series 
summation. 

The self-similar extrapolation of virial expansions for stable hard-disc and hard-sphere 
fluids in Sections V and VI, respectively, yields the equations of state with the accuracy 
not worse than the best Pad\'{e} approximants and Monte Carlo simulations. In Section VII,
the summation of virial expansions for soft-sphere fluids with power-law interaction 
potentials is considered. Again, the extrapolated compressibility factors are in perfect 
agreement with Monte Carlo calculations. Let us stress that these results are obtained being
based solely on virial expansions, without involving any fitting parameters. Section VIII
concludes.

\section{Virial expansions}    
  
Let us formulate the problem for hard spheres in spatial dimension $d$. The aim is to study
the compressibility factor
\be
\label{1}
 Z \; = \; \frac{\bt P}{\rho} \qquad 
\left( \bt \equiv \frac{1}{k_B T} , ~ \rho \equiv \frac{N}{V} \right) \;  ,
\ee
in which $\beta$ is inverse temperature, $P$, pressure, $\rho$, density, $N$, the number of
spheres, and $V$, the system volume. The virial expansion is a series in powers of density,
\be
\label{2}
Z \; = \; 1 + \sum_{n=2}^\infty B_n \rho^{n-1} \;   ,
\ee
where $B_n$ are the virial coefficients. 

It is convenient to introduce the dimensionless virial coefficients
\be
\label{3}
b_n \; \equiv \; \frac{B_n}{V_d^{n-1} } \; = \; 
\frac{B_n}{B_2^{n-1} } \; 2^{(d-1)(n-1)} \;   ,
\ee
using the notation for the second virial coefficient
\be
\label{4}
 B_2 \; = \; 2^{d-1} V_d \; , \qquad b_2 \; = \; 2^{d-1}  
\ee
and the volume of a $d$-dimensional sphere of radius $R$,
\be
\label{5}
 V_d \; = \; \frac{\pi^{d/2} R^d}{\Gm(1+d/2)} \;  .
\ee
 
As a variable, one employs the packing fraction that we denote as
\be
\label{6}
 x \; \equiv \; \rho V_d \;  .
\ee
Then the virial expansion takes the form
\be
\label{7}
 Z \; = \; 1 + \sum_{n=2}^\infty b_n x^{n-1} \;  .
\ee
 
In practice, not all virial coefficients are known, and one has only a truncated series for 
the compressibility factor that is convenient to represent as
\be
\label{8}
Z_k(x) \; = \; 1 + \sum_{n=1}^k a_n x^n \;   ,
\ee
with the coefficients
\be
\label{9}
 a_n \; \equiv \; b_{n+1} \; = \; \frac{B_{n+1}}{B_2^n} \; 2^{(d-1)n} 
\qquad 
\left( a_1 = b_2 = 2^{d-1}\right) \;  .
\ee

Sometimes, one uses as a variable the reduced density, which is related to the packing fraction
by the expression
\be
\label{10}
\frac{\rho}{\rho_d} \; = \; \frac{x}{x_d} \qquad
(x_d = \rho_d V_d) \;   ,
\ee
where $\rho_d$ is the closest packing density and $x_d$ is the closest packing fraction.

The radius of convergence of virial series, $x_{con}$ or $\rho_{con}$, is usually rather small
\cite{Lebowitz_8,Procacci_9}, which requires the virial expansions to be extrapolated to the 
physical values of the variable $x \gg x_{con}$. The stable hard-sphere fluid exists not for 
all $x$, but its region of stability is limited by a freezing packing fraction $x_f$, where 
the fluid freezes into a solid state \cite{Adler_29,Adler_30,Strandburg_31,Huerta_32,Estrada_33,
Bannerman_34}. Thus, below the radius of convergence $x_{con}$ the virial expansion as such can 
be used, while above the radius of convergence $x_{con}$ until the freezing point $x_f$, that 
is in the region
\be
\label{11}
x_{con} \; = \; x_d \; \frac{\rho_{con}}{\rho_d} \; \leq \; x \; \leq \; x_f \; = \; 
x_d \; \frac{\rho_f}{\rho_d} \;   ,
\ee
an effective summation of the virial expansion has to be invoked.  
 
For hard-core disks and spheres, the known virial coefficients \cite{Clisby_35,Wheatly_36}
yield the coefficients of expansion (\ref{8}) listed in Table 1. As is seen, the values of 
the coefficients increase with the order $k$, which hints that for the packing fraction 
$x$ above the radius of convergence the series diverge and require an effective extrapolation
from the region of small $x$.  


\section{Self-similar summation}

The method of extrapolation for expansions over asymptotically small variables to the region
of finite and even large variables has been expounded in the review articles \cite{Yukalov_27,
Yukalov_28} and in the recent papers \cite{Yukalov_37,Yukalov_38}. The main idea of the approach, 
which gave the name for it, is as follows. Suppose, we have an expansion over a small variable 
or parameter $x$. For realistic problems, one always has not an infinite series, but a 
finite-order truncated expansion, as in Eq. (\ref{8}). In order to find out what would be the 
behavior of the series for large orders $k$ and what would be the effective sum for the series, 
we try to learn what is the relation between the subsequent partial sums. Say, we wish to find 
out what is a transformation connecting the second order truncated expansion 
$Z_2(x) = a_0 + a_1 x + a_2 x^2$ and the third order expansion
$Z_3(x) = a_0 + a_1 x + a_2 x^2 + a_3 x^3$. Generally, we try to find out the relation between
$Z_k(x)$ and $Z_{k + 1}(x)$. Such a relation reveals the law of self-similarity of the series.  
This procedure is somewhat analogous to the methods of derivation of functional self-similarity
in renormalization group theory \cite{Stuckelberg_39,Wilson_40,Bogolubov_41,Yukalov_42,
Kadanoff_43}.

On the technical side, to discover self-similarity in a series, several steps are to be made.
First, to improve the series convergence, control parameters are to be introduced. This can be 
done in different ways, by including the control parameters into initial conditions, by variable 
transformations, or by series transformations. An efficient method of implanting control 
parameters is through the fractal transform, whose control parameters are defined from training 
conditions. Control parameters, in general, can be transformed into control functions defined by 
optimization conditions. Then the partial sums of the series are treated as the points of the 
trajectory of an approximation cascade. The transition from one approximation order to another
is interpreted as the temporal motion within a functional space in discrete time, where the role 
of discrete time is played by the approximation order $k$. The evolution equation of the 
approximation cascade, in the vicinity of a fixed point, is given by the self-similar relation. 
In order to pass from discrete time to continuous time, the approximation cascade is embedded 
into the approximation flow, whose trajectory passes through all points of the cascade trajectory, 
and the evolution equation satisfies the self-similarity relation. This relation can be 
reformulated as a Lie differential equation, which can then be integrated and analyzed to define 
fixed-point solutions. These solutions serve as the effective limit $Z_k^*(x)$ for the 
approximation sequence. Here and in what follows the notation for an approximant $Z_k^*(x)$, 
with a star, reminds us that this approximant plays the role of a fixed point for the 
trajectory of an approximation cascade. This effective limit, representing a fixed point, 
is termed a self-similar approximant of order $k$. The latter can be represented in the form 
of self-similar factor approximants \cite{Yukalov_44,Gluzman_45,Yukalov_46,Yukalov_2009}. 
More details can be found in reviews \cite{Yukalov_27,Yukalov_28} and in the recent papers 
\cite{Yukalov_37,Yukalov_38}. 

Of course, there is no need to repeat each time all those mathematical manipulations described 
above. They are sketched here just to give the reader the idea of why the approximations are 
termed self-similar and what is their justification. Actually, all we need is the outcome of 
this procedure that results in the self-similar factor approximants 
\be
\label{12}
 Z_k^*(x) \; = \; \prod_{i=1}^{N_k} ( 1 + A_i x)^{n_i} \;  ,
\ee
where
\begin{eqnarray}
\label{13}
 N_k \; = \; \left\{ \begin{array}{rr}
k/2 \; , ~ & ~ k = 2,4,\ldots \\
(k+1)/2 \; , ~ & ~ k = 3,5, \ldots
\end{array} 
\right.
\end{eqnarray}
and $A_i$ and $n_i$ are control parameters defined by training conditions. 

The natural condition that has to be imposed on a self-similar approximant is the equivalence 
of the asymptotic, as $x \ra 0$, expression of the approximant and of the initial virial 
expansion,
\be
\label{A1}
Z_k^*(x) \; \simeq \; Z_k(x) \qquad (x \ra 0) \;   .
\ee
The approximant asymptotic expression is
\be
\label{A2}
 Z_k^*(x) \simeq \; \sum_{n=0}^k \frac{1}{n!} \; \left[ \;
\frac{\prt^n Z_k^*(x)}{\prt x^n} \; \right]_{x=0} \; x^n 
\qquad (x \ra 0 ) \;  .
\ee
Equating the similar terms in the above asymptotic expression and in the given virial expansion
yields the training condition
\be
\label{14}
\lim_{x\ra 0} \; \frac{1}{n!}\; \frac{d^n}{dx^n} \; Z_k^*(x) \; = \; a_n \;  .
\ee
Using the scaling properties of approximants (\ref{12}), one of the parameters $A_i$, for
odd orders $k$ can be set to one \cite{Yukalov_27,Yukalov_28,Yukalov_38}.  

As is seen, the form of the self-similar approximants (\ref{12}) is more general than that of
Pad\'{e} approximants, which can be considered as a very particular case of the self-similar 
factor approximants, since the latter can be reduced to the former by setting $n_i = \pm 1$. 
The presence of nontrivial powers $n_i$ allows for the better approximation of both rational 
and irrational functions, contrary to the rational Pad\'{e} approximants.  

When dealing with self-similar factor approximants of high-orders, it is more convenient to 
use, instead of Eq. (\ref{14}), the equality for the logarithms
\be
\label{A3}
 \ln Z_k^*(x) \simeq \ln Z_k(x) \qquad ( x\ra 0) \;  .
\ee
Then the training conditions for the control parameters $A_i$ and $n_i$ read as
\be
\label{A4}
 \sum_{i=1}^{N_k} n_i\;  A_i^m \; = \; \Lbd_m \qquad (m=1,2,\ldots,k) \;  ,
\ee
where
$$
 \Lbd_m \; = \; \frac{(-1)^{m-1}}{(m-1)!} \; \lim_{x\ra 0} \;
\frac{d^m}{d x^m} \; \ln Z_m(x) \;  .
$$
  
If $A_i > 0$, the self-similar approximant has no singularities for finite $x$. However,
if there exists $A_m$ such that
\be
\label{15}
A_m \; \equiv \; \min_i A_i \; < \; 0 \; , \qquad n_m \; < \; 0 \;   ,
\ee
then there appears a singularity in whose vicinity the approximant behaves as
\be
\label{16}
Z_k^*(x) \; \simeq \; \frac{C_k}{(x_c-x)^\al} \;   ,
\ee
with the critical parameters
\be
\label{17}
x_c \; = \; \frac{1}{|\; A_m\;|} \; , \qquad \al \; = \; |\; n_m \; | \; ,
\qquad
C_k \; = \; Z_k^*(x_c) \; ( 1 + A_m x_c)^\al \; .
\ee

When the approximants $Z_k^*(x)$ exhibit numerical convergence to the form (\ref{16}), this
means that the singularity is motivated by the physics of the problem, although it may occur
on a metastable branch. In the case of hard spheres, the appearance of the pole $x_c$ is 
caused by the existence of the closest packing, hence $x_c$ should be close to $x_d$, however
not necessarily coinciding with the latter.

\section{Hard-rod fluid}

As a simple and instructive example of employing self-similar approximants, let us consider 
a system of hard rods. For the spatial dimensionality $d=1$, we have $V_1 = 2R$ and the 
packing fraction $x = 2 \rho R$. The closest packing density is $\rho_1 = 1/ 2R$, and the 
closest packing fraction is $x_1 = 1$. The virial coefficients are $a_n = b_{n+1} = 1$. 
Thus, the virial expansion becomes
\be
\label{18}
Z_k(x) \; = \; 1 + \sum_{n=1}^k x^n \;  .
\ee

Let us make impression that we do not know how geometric series are summed. Instead, in order 
that the reader could easily understand how self-similar factor approximants are constructed, 
let us explain the technical details on the simple case of hard rods. 

Starting with the second-order virial expansion 
\be
\label{A5}
Z_2(x) = 1 + x + x^2 \;   ,
\ee
we have the factor approximant
\be
\label{A6}
Z_2^*(x) \; = \; ( 1 + A_1 x )^{n_1} \;   .
\ee
Expanding the above expression in powers of $x$ results in
\be
\label{A7}
 Z_2^*(x) \; \simeq \; 
1 + n_1 A_1 x + \frac{1}{2} \; n_1 \; ( n_1 -1 )\; A_1^2 \; x^2 \;  .
\ee
Equating the similar terms of (\ref{A5}) and (\ref{A7}) gives the training conditions
$$
 n_1 \; A_1 \; = \; 1 \; , \qquad 
\frac{1}{2} \; n_1 \; ( n_1 -1 ) \; A_1^2 \; = \; 1 \;  ,
$$
whose unique solutions are $A_1 = -1$ and $n_1 = -1$. Hence $Z_2^*(x) = 1/(1-x)$. 

If we prefer to compare the logarithms (\ref{A3}) and to use conditions (\ref{A4}), we 
have to compare the expansions
\be
\label{A8}
 \ln Z_2(x) \; \simeq \; x + \frac{1}{2} \; x^2 \; ,
\qquad
 \ln Z_2^*(x) \; \simeq \;n_1 \; A_1 \; x -\;  \frac{1}{2} \; n_1 \; A_1^2 \; x^2 
\qquad ( x \ra 0) \; ,
\ee
which gives the training conditions 
$$
n_1 \; A_1 \; = \; 1 \; , \qquad 
-\; \frac{1}{2} \; n_1 \; A_1^2 \; = \; \frac{1}{2}   .
$$
The unique solutions of the latter are again $A_1 = -1$ and $n_1 = -1$, hence again we have
$Z_2^*(x) = 1/(1-x)$.

The third-order virial expansion 
\be
\label{A9}
 Z_3(x) \; = \; 1 + x + x^2 + x^3  
\ee
lead to the self-similar factor approximant
\be
\label{A10}
Z_3^*(x) \; = \;  (1 + x)^{n_1} ( 1 + A_2 x )^{n_2} \;  .
\ee
For $x\ra 0$ the corresponding logarithms give
$$
\ln Z_3(x) \; \simeq \; x + \frac{1}{2} \; x^2 + \frac{1}{3} \; x^3 \; ,
$$
\be
\label{A11}
\ln Z_3^*(x) \; \simeq \; ( n_1 + n_2 \; A_2) \; x - \;
\frac{1}{2} \; \left( n_1 + n_2 A_2^2 \right) \; x^2  +
\frac{1}{3} \; \left( n_1 + n_2 A_2^3 \right) \; x^3 \; .
\ee
Comparing these expressions, we get the training conditions   
$$
 n_1 + n_2 \; A_2 \; = \; 1 \; , \qquad
-\; \frac{1}{2} \; \left( n_1 + n_2 \; A_2^2 \right) \; = \; \frac{1}{2} \; ,
\qquad
\frac{1}{3} \; \left( n_1 + n_2 \; A_2^3 \right) \; = \; \frac{1}{3} \;  ,
$$
whose unique solutions are $n_1 = 0$, $n_2 = -1$, and $A_2 = -1$. Therefore, we again have
the approximant $Z_3^* = 1/(1-x)$. Continuing in that way, we obtain
\be
\label{19}
Z_k^*(x) \; = \; \frac{1}{1-x} \; \qquad ( k\geq 2) \;   ,
\ee
which is the exact expression for sum (\ref{18}). This example shows that in some cases the 
self-similar approximants can reconstruct the sought functions exactly.  

Moreover, any real function of the form 
$$
f_{k_M}(x) \; \equiv \; \prod_{j=1}^M P_{m_j}^{\al_j}(x) \;    ,
$$
being a product of polynomials $P_{m_j}(x)$ in powers $\alpha_j$, can be exactly reconstructed 
by means of the factor approximants of order 
$$
k_M \; = \; M + \sum_{j=1}^M m_j \;   .
$$

\section{Hard-disc fluid}

In spatial dimensionality $d = 2$, the components are hard discs of radius $R$. Then 
$V_2 = \pi R^2$ and $B_2 = 2 V_2 = 2 \pi R^2$. The packing fraction is
\be
\label{20}
x \; = \; \rho V_2 \; = \; \pi \rho R^2 \;   .
\ee
For the closest packing density, one has $\rho_2 R^2 = 1/2 \sqrt{3} = 0.2887$, and the closest
packing fraction is
\be
\label{21}
x_2 \; = \; \frac{\pi}{2\sqrt{3}} \; = \; 0.9069 \;   .
\ee

The estimates for the radius of convergence \cite{Lebowitz_8} and the freezing density 
\cite{Clisby_35} are defined by the relative ratios
$$
\frac{\rho_{con}}{\rho_2} \; = \; 0.0399 \; , \qquad 
\frac{\rho_f}{\rho_2} \; = \; 0.78 \;   .
$$
This gives for the related packing fractions
\be
\label{22}
x_{con} \; = \; 0.0362 \; , \qquad x_f \; = \; 0.7074 \;   .
\ee
 
The factor approximants, constructed from expansion (\ref{8}), with the coefficients from 
Table 1, have the form (\ref{12}). For illustration, we present several first approximants
explicitly. In the second order, we have
$$
Z_2^*(x) \; = \; \frac{1}{(1-1.1280x)^{1.7730} } \;   ,
$$
in third order,
$$
Z_3^*(x) \; = \; \frac{1}{ (1-1.0013x)^{2.1239} (1+x)^{0.1267} } \;    ,
$$
in fourth order,
$$
Z_4^*(x) \; = \; \frac{(1-0.0444x)^{13.4304}}{(1-0.8791x)^{2.9533}} \;  ,
$$
in fifth order,
$$
Z_5^*(x) \; = \;
\frac{(1+x)^{0.0761}}{ (1-0.9154x)^{2.5907} (1+0.3599)^{1.2440} } \;   ,
$$
and so on. 

In the vicinity of the singularity point $x_c$, the behavior of the approximants is as given 
by Eq. (\ref{16}), with the critical parameters listed in Table 2. The singularity point 
is close to $1$, in agreement with the pole usually accepted in the empirical equations 
of state for hard-disc fluid. The critical exponent $\alpha$ is also close to $2$ used in 
the phenomenological equations of state. Note that in our case the pole and the critical 
exponent are not fitted but naturally appear in the frame of the approximation.


The overall behavior of the self-similar approximation $Z^*_k(x)$ for the compressibility 
factor in the stable region of the packing fraction $x \in [0, x_f]$ can be compared with 
the phenomenological equation of state 
\be
\label{23}
 Z_L(x) \; = \; \frac{1+\frac{1}{8} x^2 + \frac{1}{18} x^3 - \frac{4}{21} x^4}{(1-x)^2} 
\qquad
(d = 2)   
\ee
suggested by Liu \cite{Liu_47} for the same stable region of the hard-disk fluid. In this 
region, the Liu compressibility factor (\ref{23}) is obtained by simultaneously fitting the 
compressibility factor to molecular-dynamics computer simulations \cite{Kolafa_48,Luban_49}, 
as well as to all available virial coefficients. The factor (\ref{23}) describes the equation 
of state for hard disks with the highest possible accuracy referring to the reproduction of 
the compressibility data with the accuracy comparable to that of the highest-level Pad\'{e} 
approximations. The self-similar approximants $Z_k^*(x)$ are very close to the Liu equation 
of state, with their high orders practically coinciding with Eq. (\ref{23}). Since the lines 
for $Z_k^*(x)$ and $Z_L(x)$ are not distinguishable between each other, we show in Fig. 1 the 
relative difference 
\be
\label{24}
 \epsilon_k(x) \; = \; \frac{Z_k^*(x) - Z_L(x)}{ Z_L(x) } \times 100\% \; ,
\ee
measured in percent, as a function of the packing fraction in the stable region. We show the
difference (\ref{24}) for the self-similar approximants $Z_6^*(x)$, $Z_7^*(x)$ and $Z_9^*(x)$,
while the approximant $Z_8^*(x)$ is not distinguishable from $Z_9^*(x)$. The parameters $A_i$
and $n_i$ for these approximants are given in Appendix A. 


The largest difference, as is expected, occurs at the freezing point, where, nevertheless, 
it is rather small, being $\epsilon_6(x_f) = 0.117\%$, $\epsilon_7(x_f) = 0.019\%$, and 
$\epsilon_8(x_f) = \epsilon_9(x_f) = 0.040\%$. 

It is important to stress that each self-similar approximant $Z_k^*(x)$ of order $k$, by 
construction, reproduces exactly the virial expansion of order $k+1$, with exact virial 
coefficients $a_n \equiv b_{n+1}$ for $n = 1,2,\ldots, k$. Moreover, it predicts sufficiently 
accurately the higher-order virial coefficients above the order $k+1$. Thus, the second-order 
compressibility factor $Z_2^*(x)$, being expanded in powers of $x$, 
$$
Z_2^*(x) \; \simeq Z_2(x) + 4.4377 x^3 + 5.9731 x^4 \;  ,
$$
gives exactly the virial expansion $Z_2(x)$, with exact virial coefficients $a_1 = b_2$ and 
$a_2 = b_3$, and also predicts the higher coefficients, for example $a_3 = b_4$ and 
$a_4 = b_5$ that should be compared with the values in Table 1. Similarly, expanding 
$Z_k^*(x)$ in $x$, we exactly reproduce the virial expansions $Z_k(x)$, with all exact virial 
coefficients up to the order $k+1$, and get a prediction for the higher-order virial 
coefficients, as is shown in the expansions
$$
Z_3^*(x) \; \simeq Z_3(x) + 5.4613 x^4 + 6.6671 x^5 \; , 
$$
$$
Z_4^*(x) \; \simeq Z_4(x) + 6.3263 x^5 + 7.2025 x^6 \; , 
$$
$$
Z_5^*(x) \; \simeq Z_5(x) + 7.2903 x^6 + 8.1353 x^7 \;   ,
$$
$$
Z_6^*(x) \; \simeq Z_6(x)  + 8.3177 x^7 + 9.2676 x^8\;   ,
$$
$$
Z_7^*(x) \; \simeq Z_7(x) + 9.2732  x^8 + 10.2182 x^9 \;   ,
$$
$$
Z_8^*(x) \; \simeq Z_8(x) + 10.2158 x^9 \; .
$$
The accuracy of predictions can be compared with the known virial coefficients in Table 1. 
For instance, in the expansion of $Z_8^*(x)$, we get exactly the virial expansion $Z_8(x)$
and predict the coefficients $a_9 = b_{10} = 10.2158$, which is very close to the known 
coefficient $a_{9} = b_{10} = 10.2163$, as is given in Table 1.

\section{Hard-sphere fluid}

For the hard-sphere fluid (d=3), with radius $R$, we have $V_3 = (4 \pi/3) R^3$ and
$B_2 = 4 V_3 = (16 \pi/3) R^3$. Therefore the packing fraction is
\be
\label{25}
 x \; = \; \rho V_3 \; = \; \frac{4\pi}{3} \; \rho R^3 \;  .
\ee
For the closest packing density, we have $\rho_3 R^3 = 1/4\sqrt{2} = 0.1768$, hence the closest
packing fraction is
\be
\label{26}
 x_3 \; = \; \frac{\pi}{3\sqrt{2}} \; = \; 0.7405 \;  .
\ee  
The radius of convergence \cite{Lebowitz_8} and the freezing point with respect to density 
\cite{Clisby_35} can be estimated by the ratios
$$
\frac{\rho_{con}}{\rho_3} \; = \; 0.0244 \; , \qquad
\frac{\rho_f}{\rho_3} \; = \; 0.66 \;   , 
$$
so that the corresponding packing fractions are
\be
\label{27}
 x_{con} \; = \; 0.0181 \; , \qquad x_f \; = \; 0.4887 \;  .
\ee

Composing, on the basis of the virial expansions (\ref{8}), the self-similar factor approximants 
$Z_k^*(x)$, we find the critical behavior (\ref{16}) with the critical parameters listed in 
Table 3. The behavior of the compressibility factors $Z_k^*(x)$ in the whole stable region of 
the packing fraction $x$ can be compared with the Liu \cite{Liu_47} phenomenological equation 
of state
\be
\label{28}
 Z_L(x) \; = \;
\frac{1+x+x^2-\frac{8}{13} x^3 - x^4 + \frac{1}{2} x^5}{(1-x)^3}   
\ee
obtained by simultaneously fitting the virial coefficients up to $11$-th order 
\cite{Wheatly_36,Shultz_50} and the available compressibility data from molecular dynamics 
simulations \cite{Pieprzyk_51}. This equation (\ref{28}), yielding accurate results for the
compressibility factor, also very well agrees with the isothermal compressibility found in
Monte Carlo simulations \cite{Grigoriev_52}.   


The high-order self-similar approximants practically coincide with the equation of state 
(\ref{28}), therefore Figure 2 shows the relative difference (\ref{24}) in percent between 
self-similar approximants $Z_k^*(x)$ and equation (\ref{28}). The difference for the 
approximants $Z_8^*(x)$, $Z_9^*(x)$, and $Z_{10}^*(x)$ is shown. The parameters of these 
approximants are given in Appendix B. 


The maximal difference occurs at the freezing point, where it is also not so high: 
$\epsilon_8(x_f) = 0.064 \%$, $\epsilon_9(x_f) = 0.072 \%$, and 
$\epsilon_{10}(x_f) = 0.068 \%$.  

By expanding the self-similar factor approximants $Z_k^*(x)$ of order $k$ in powers of $x$ 
we obtain the virial expansion of order $k$ with all exact virial coefficients up to 
$a_k = b_{k+1}$, that is, exactly reproducing all virial coefficients up to $B_{k+1}$, and 
the following terms predict the higher virial coefficients. Thus, expanding the approximant 
$Z_k^*(x)$, we get the exact virial expansion $Z_k(x)$ and the following terms predicting 
the higher virial coefficients:
$$
Z_2^*(x) \; \simeq \; Z_2(x) + 20 x^3 + 35 x^4 \; 
$$
$$
Z_3^*(x) \; \simeq \; Z_3(x) + 27.9847 x^4 + 36.6223 x^5 \; ,
$$
$$
Z_4^*(x) \; \simeq \; Z_4(x) + 37.3087 x^5 + 44.7192 x^6 \;   ,
$$
$$
Z_5^*(x) \; \simeq \; Z_5(x) + 76.5668 x^6 + 387.6802 x^7 \;   ,
$$
$$
Z_6^*(x) \; \simeq \; Z_6(x) + 67.9833 x^7 + 83.5375 x^8 \;   ,
$$
$$
Z_7^*(x) \; \simeq \; Z_7(x) + 85.3917 x^8 + 104.3735 x^9 \;   ,
$$
$$
Z_8^*(x) \; \simeq \; Z_8(x) + 105.7174 x^9 + 127.8835 x^{10} \;   ,
$$
$$
Z_9^*(x) \; \simeq \; Z_9(x) + 128.0845 x^{10} + 153.0054 x^{11} \;   ,
$$
$$
Z_{10}^*(x) \; \simeq \; Z_{10}(x) + 153.1035 x^{11} \;   ,
$$
The higher the order of the approximant $Z_k^*(x)$, the better it predicts the following virial
coefficients. Of the most interest is the prediction of the twelfth virial coefficient $B_{12}$, 
which is defined by the coefficient $a_{11} = b_{12}$. The known estimates of the coefficient 
$b_{12}$ give the values: $149$ in \cite{Liu_47}, $153$ in \cite{Hu_18,Tian_20,Maestre_21,
Clisby_35}, and $154$ in \cite{Lekkerkerker_53}. Our prediction, as is seen from the expansion 
for $Z_{10}^*(x)$, equals $153$ that coincides with the value predicted by many authors 
\cite{Hu_18,Tian_20,Maestre_21,Clisby_35}.

\section{Soft-sphere potentials}

Virial expansions of hard-core fluids, treated above, do not depend on temperature. This
is not so for soft-sphere potentials. However the method of self-similar summation can be
applied for soft-sphere potentials as well. For illustration, we consider three-dimensional 
fluids with the repulsive power-law interaction potential
\be
\label{B1}
 \Phi(r) \; = \; \ep \; \left( \frac{\sgm}{r} \right)^p \;  ,
\ee
where the positive parameters $\varepsilon$, $\sigma$, and $p$ characterize the typical 
energy, diameter, and effective hardness of the spheres, respectively. Fortunately, it is
possible \cite{Barlow_2012} to introduce the effective virial coefficients   
\be
\label{B2}
 \overline B_n \; = \; B_n \; \left[ \;
(  \bt \ep )^{3/p} \; \sgm^3 \; \right]^{1-n}  
\ee
and the expansion variable
\be
\label{B3}
  y \; \equiv \; \rho \; \sgm^3 (\bt \ep)^{3/p} \;  ,
\ee
playing the role of an effective density, so that the virial expansion reduces to
\be
\label{B4}
 Z \; = \; 1 + \sum_{n=2}^\infty \overline B_n y^{n-1} \; .
\ee

With the renotation
\be
\label{B5}
 c_n \; \equiv \; B_{n+1} \; , \qquad
c_1 \; = \; \overline B_2 \; = \; 
\frac{2\pi}{3} \; \Gm\left( 1 - \; \frac{3}{p} \right) \;  ,
\ee
we come to the standard form of virial expansion for the compressibility factor
\be
\label{B6}  
Z_k(y) = 1 + \sum_{n=1}^k c_n y^n \;   .
\ee

The reduced virial coefficients for different powers $p$ have been summarized by Barlow 
et al. \cite{Barlow_2012}. For instance for the powers $p=6$ and $p=12$, that are often 
met in applications, the effective virial coefficients in our notation, $c_n$, are listed 
in Table IV. Self-similar factor approximants $Z_k^*(y)$, defined as in (\ref{12}), 
represent the fluid equation of states for the variable in the range $0 < y \leq y_f$, where
$y_f$ is a reduced freezing  density. In the case of $p = 6$, we have \cite{Tan_2011} the 
reduced freezing density $y_f = 2.3$ and for $p=12$, the freezing point is $y_f = 1.14$.  
The parameters of the self-similar factor approximants $Z_6^*$ and $Z_7^*$ are given in 
Appendix C for $p=6$ and in Appendix D for $p=12$. The numerical convergence of the 
approximants is shown in Table 5 for $p=6$ and in Table 6 for $p=12$. Our results are in 
very good agreement with the Monte Carlo simulations \cite{Tan_2011}. The soft-sphere fluids
with other power-law potentials are also well described by using self-similar summation.     

Although the variable $y \in [0,y_f]$ is limited by the freezing point $y_f$, it can be 
useful to find out the limiting behavior at $y \ra \infty$. The knowledge of this asymptotic
tendency can help improving the accuracy of the equation of state in the physical region of 
$y$, as is discussed by Barlow et al. \cite{Barlow_2012}. From the expression of the 
self-similar approximant
\be
\label{B7}
Z_k^*(y) \; = \; \prod_{i=1}^{N_k} ( 1 + A_i y)^{n_i} \;   ,   
\ee
it follows that at large $y$, we have
\be
\label{B8}
 Z_k^*(y) \; \simeq \; B_k y^{\nu_k^*} \qquad ( y \ra\infty) \;  ,
\ee
where $\nu_k^*$ is a large-variable exponent of the self-similar approximant,
\be
\label{B9}
 B_k  \; = \; \prod_{i=1}^{N_k} A_i^{n_i} \; , 
\qquad 
\nu_k^* \; = \; \sum_{i=1}^{N_k} \nu_i \;  .
\ee

For the considered powers $p$, we find, in the approximations $k=6$ and $k=7$, the 
asymptotic exponents
$$
\nu_6^* \; = \; 2.064 \; , \qquad \nu_7^* \; = \; 1.989 \qquad ( p = 6) \; ,
$$
$$
\nu_6^* \; = \; 5.059 \; , \qquad \nu_7^* \; = \; 4.784 \qquad ( p = 12) \; .
$$
Taking the half sums of the amplitudes $B_6$ and $B_7$ and of the exponents $\nu_6^*$ and 
$\nu_7^*$, we come to the asymptotic behavior
$$
Z^*(y) \; \simeq \; 6.54 y^{2.03} \qquad ( p = 6) \; ,
$$
\be
\label{B10}
Z^*(y) \; \simeq \; 1.19 y^{4.92} \qquad ( p = 12) \;   .
\ee
This is close to the estimate $p/3$, that sometimes is assumed \cite{Barlow_2012}.

\section{Conclusion}

A new approach to the summation of virial expansions is suggested. The approach follows the 
idea of self-similar approximation theory, which is based on finding the similarity between 
the subsequent approximations in the region of a small variable and extrapolating the resulting 
form to the whole region of the variable of interest. Then, analyzing the similarity between 
the virial expansions of different orders $k$ for the compressibility factor $Z_k(x)$, in powers 
of a small parameter, such as packing fraction, $x \ra 0$, it is possible to find a relation 
between the subsequent expansion orders $Z_k(x)$ and $Z_{k+1}(x)$ and extrapolate the found 
similarity law to the whole region of $x$, where the system remains stable, thus obtaining 
the self-similar approximant $Z^*_k(x)$ valid for all $x$ in the stability region. The method 
is regular. Its validity is confirmed by the sequence of self-similar approximants 
demonstrating numerically close results.       

The approach is illustrated by summing the virial expansions for hard-disk, hard-sphere, and
soft-sphere fluids. In the case of the hard-rod fluid, the self-similar summation yields the 
exact result. In all considered cases, for hard-disk, hard-sphere, and soft-sphere fluids, 
the obtained self-similar approximants for the compressibility factor, using no fitting 
parameters, are in perfect agreement with the phenomenological equations of state fitted to 
numerical compressibility data, as well with direct Monte Carlo simulations. The approximants 
of the $k$-th order exactly reproduce the known virial coefficients up to $B_{k+1}$ and predict 
well the higher-order virial coefficients.

Summarizing, the advantage of the method of self-similar approximants is in the following:
\begin{itemize}
\item
It is general, being valid for any given expansion. The results are reliable when the sequence
of self-similar approximants exhibits numerical convergence. 
 
\item
It is regular, being convergent for convergent series.

\item
It is uniquely defined, with the parameters prescribed by training conditions.

\item
It gives an accurate extrapolation of asymptotic series to the region of the variables
above the range of convergence.

\item
Taking account of a singularity, related to the closest packing, is done automatically,
without the need of its artificial separation. 

\item
The extrapolated form exactly reproduces all virial coefficients contained in the initial
expansion. 

\item
It predicts the virial coefficients of higher orders, above those used in the initial 
expansion.   

\item
It is based on a mathematically clear idea of discovering similarity between the 
subsequent terms of an expansion.
\end{itemize}  

This approach can be used for constructing explicit approximations for the equations of state 
of systems with repulsive interaction potentials. The sole thing that is required for this 
purpose is to have a virial expansion at small density or packing fraction, with a sufficient 
number of terms allowing for the observation of numerical convergence of self-similar 
approximants. 

The application of the method to systems with interaction potentials containing attractive 
terms, such as the Lennard-Jones potential, requires a separate investigation. Virial 
expansions for these potentials have rather different form depending on whether temperature
pertains to subcritical or supercritical region \cite{Kolafa_1994,Singh_10,Barlow_2014,
Barlow_2015,Feng_11,Elliott_2019,Gottschalk_2019}. Virial expansions in the supercritical 
region allow for self-similar summation. However the virial coefficients in the subcritical 
region are all negative, which makes problems for the direct application of summation methods. 
From the mathematical point of view, the standard virial expansion for systems with attractive 
interactions, below critical temperature, does not seem to have the structure allowing for 
straightforward summation, but some rearrangement of the series is necessary. Additional
information on the behavior of the system in the high-density limit \cite{Barlow_2012} or 
near the critical point \cite{Barlow_2014,Barlow_2015} may be necessary for improving the
convergence of approximants. These delicate problems need additional consideration.   

\section*{Declarations}

{\parindent=0pt
{\bf Author contributions}: All authors contributed to the study conception and design. 
Material preparation, data collection and analysis were performed by V.I. Yukalov and 
E.P. Yukalova. Numerical calculations were done by E.P. Yukalova. The first draft of the 
manuscript was written by V.I. Yukalov and all authors commented on previous versions 
of the manuscript. All authors read and approved the final manuscript.

\vskip 2mm
{\bf Funding}: No funding was received to assist with the preparation of this manuscript.

\vskip 2mm
{\bf Financial interests}: The authors have no relevant financial or non-financial 
interests to disclose. 

\vskip 2mm
{\bf Competing interests}: The authors have no competing interests to declare that are 
relevant to the content of this article. }

\newpage

\section*{Appendix A. Hard disks}

The relative difference (\ref{24}) in percent between the self-similar factor approximants 
$Z_6^*(x)$, $Z_7^*(x)$ and $Z_9^*(x)$ for the compressibility factor, and the phenomenological 
equation of state (\ref{23}) is shown in Fig. 1. 

Recall that here and in what follows the notation $Z_k^*(x)$ implies that this is a 
self-similar approximant of order $k$ for the compressibility factor. The parameters 
$A_i$ and $n_i$ of the approximants can be real-valued or complex-valued. In the latter case,
they enter the approximant in complex-conjucate pairs, so that the approximant yields real
values, as it must. 

Below, in Appendices A,B,C, and D, complex-conjugate values for $A_i$ and $n_i$ are 
denoted as $A_i^*$ and $n_i^*$.  
  
The parameters for $Z_k^*(x)$ are as follows. For $Z_6^*(x)$, the parameters are:
$$
A_1 =  -1.003351 \; , \qquad A_2 = -0.284165 + i\; 0.570994 \; , 
\qquad A_3 = A_2^*  \; ,
$$
$$
n_1 = -1.816418 \; , \qquad  n_2 = 0.401376 - i \; 0.355178 \; , \qquad
n_3 = n_2^* \; .
$$
For $Z_7^*(x)$, the parameters are:
$$
A_1 = 1 \; , \qquad
A_2 =  -0.295714 - i\; 0.555743  \; , \qquad A_3 = A_2^* \; , 
\qquad
A_4 =  -1.007366 \; ,
$$
$$
n_1 = 0.000634 \; , \qquad  n_2 =  0.413914 + i \; 0.405165 \; , \qquad
n_3 = n_2^* \; , \qquad n_4 = -1.780716  \; .
$$
And for $Z_9^*(x)$, the parameters are: 
$$
A_1 = 1 \; , \qquad 
A_2 = -1.006335  \; , \qquad 
A_3 = - 0.293207 + i \; 0.561130  \; ,
$$
$$
A_4 = A_3^* \; , \qquad 
A_5 = 2.41033934 \; ,
$$
$$
n_1 = 0.000253 \; , \qquad 
n_2 = - 1.790402\; , \qquad 
n_3 = 0.407954 - i\; 0.389592 \; ,
$$
$$
n_4 = n_3^* \; , \qquad 
n_5 = 3.694482\cdot 10^{-6} \; .
$$

\section*{Appendix B. Hard spheres}

The relative difference (\ref{24}) between the self-similar approximants $Z_8^*(x)$, 
$Z_9^*(x)$, and $Z_{10}^*(x)$ and the phenomenological equation of state (\ref{28}) is shown
in Fig. 2. The parameters for the self-similar approximants are as follows.

For $Z_8^*(x)$, the parameters are: 
$$
A_1 =  0.066044 \; , \qquad 
A_2 =  -1.071335 \; , \qquad
A_3 =  0.513505 + i \; 1.309205\; \qquad
A_4 =  A_3^* \; , 
$$
$$
n_1 =  12.959704 \; , \qquad 
n_2 =  -2.092858 \; , \qquad
n_3 = 0.652455 - i\; 0.088547 \; \qquad
n_4 = n_3^* \; .
$$
For $Z_9^*(x)$, the parameters are: 
$$
A_1 = 1 \; , \qquad 
A_2 = -1.077386  \; , \qquad 
A_3 =  -0.081600 \; ,
$$
$$
A_4 = 0.506784 - i\; 1.312598 \; , \qquad 
A_5 = A_4^* \; ,
$$
$$
n_1 = 0.033789 \; , \qquad 
n_2 = -2.032993\; , \qquad 
n_3 =  -10.249876 \; ,
$$
$$
n_4 =  0.646037 + i\; 0.108450 \; , \qquad 
n_5 =  n_4^* \; ,
$$
And for $Z_{10}^*(x)$, the parameters are: 
$$
A_1 =  0.032194 \; , \qquad 
A_2 =   -1.073480 \; , \qquad 
A_3 =  0.5130676 - i\; 1.311983 \; ,
$$
$$
A_4 =  A_3^* \; , \qquad 
A_5 =  4.145197 \; ,
$$
$$
n_1 = 27.112178  \; , \qquad 
n_2 = -2.073162 \; , \qquad 
n_3 = 0.646069 + i\; 0.090964  \; ,
$$
$$
n_4 = n_3^*  \; , \qquad 
n_5 =  1.338641 \cdot 10^{-6} \; .
$$

\section*{Appendix C. Soft spheres ($p = 6$ )} 

Parameters of the self-similar factor approximants $Z_k^*(y)$ for the compressibility factor
of soft-sphere fluid with the power-law potential of power $p=6$ are as follows. 

For $Z_6^*(y)$, the parameters are:

$$
A_1 = 2.26235 \; , \qquad A_2 = 1.42666 - i\; 1.43529 \; , 
\qquad A_3 = A_2^* \; ,
$$
$$
n_1 = 0.195421 \; , \qquad n_2 =  0.934498 + i\; 0.210303 \; , 
\qquad n_3 = n_2^* \;  .
$$

And for $Z_7^*(y)$, the parameters are:  
$$
A_1 = 1 \; , \qquad
A_2 = 1.3282 \; , \qquad A_3 = 1.47451 + i\; 1.48632 \; , \qquad
A_4 = A_3^* \; ,
$$
$$
n_1 = - 1.79975 \; , \qquad n_2 = 2.16476 \; , \qquad
n_3 = 0.811981 - i\; 0.0814728  \; ,
\qquad n_4 = n_3^* \; .
$$

\section*{Appendix D. Soft spheres ($p = 12$)} 

Parameters of the self-similar factor approximants $Z_k^*(y)$ for the compressibility factor
of soft-sphere fluid with the power-law potential of power $p=12$ are as follows. 

For $Z_6^*(y)$, the parameters are:

$$
A_1 = 0.752884 \; , \qquad A_2 = 0.296313 + i\; 0.736677 \; ,
\qquad A_3 = A_2^* \; ,
$$
$$
 n_1 = 0.896191 \; , \qquad n_2 = 2.08114 - i\; 0.446899 \; ,
\qquad n_3 = n_2^* \; .
$$

And for $Z_7^*(y)$, the parameters are:  
$$
A_1 = 1 \; , \qquad
A_2 = 4.69293 \; , \qquad A_3 = 0.283084 + i\; 0.715014 \; ,
\qquad A_4 = A_3^* \; ,
$$
$$
n_1 = 0.371976 \; , \qquad n_2 = - 0.0000133367 \; , 
\qquad n_3 = 2.20617 - i\; 0.661199 \; , \qquad n_4 = n_3^* \; .
$$

\newpage

\begin{center}
{\large{\bf Table 1}}
\end{center}

\begin{table}
\tbl{Virial-expansion coefficients for hard-disk and hard-sphere fluids.}    
{\begin{tabular}{lll} \toprule
$n$~~~~~~  & $a_n ~ (d=2)$~~~~~~  & $a_n ~ (d=3)$   \\ \midrule
1    &   2             &  4   \\
2    &   3.1280178     &  10      \\
3    &   4.2578545     &  18.3647684     \\
4    &   5.3368966     &  28.22451    \\
5    &   6.36296       &  39.81515  \\ 
6    &   7.35186       &  53.3444  \\
7    &   8.31910       &  68.538   \\
8    &   9.27215       &  85.813  \\
9    &   10.2163       &  105.78  \\ 
10   &                 &  127.9    \\ \bottomrule
\end{tabular} }
\label{Table 1} 
\end{table}

\newpage

\begin{center}
{\large{\bf Table 2}}
\end{center}

\begin{table}
\tbl{Critical characteristics of hard-disk fluids. Critical packing 
fraction $x_c$, critical exponent $\al$, and coefficient $C_k$, for self-similar approximants 
of order $k$.}
{\begin{tabular}{llll} \toprule
$k$~~~~~~~~  &    $x_c$~~~~~~~~    &   $\al$~~~~~~~~   &   $C_k$ \\ \midrule
2    &   0.8865    &   1.7730  &   0.8077  \\
3    &   0.9987    &   2.1239  &   0.9135   \\
4    &   1.1375   &   2.9533   &   0.7294   \\
5    &   1.0924    &   2.5907  &   0.8805  \\ 
6    &   0.9967    &   1.8164  &   1.4911  \\
7    &   0.9927    &   1.7807  &   1.5435  \\
8    &   0.9939    &   1.7924  &   1.5245  \\
9    &   0.9937    &   1.7904  &   1.5279  \\ \bottomrule
\end{tabular}
}
\label{Table 2}
\end{table}

\newpage

\begin{center}
{\large{\bf Table 3}}
\end{center}

\begin{table}
\tbl{
Critical characteristics of hard-sphere fluids. Critical packing fraction $x_c$, 
critical exponent $\al$, and coefficient $C_k$, for self-similar approximants of order $k$.}
{\begin{tabular}{llll} \toprule
$k$~~~~~~  &    $x_c$~~~~~~    &   $\al$~~~~~~   &   $C_k$   \\ \midrule
6    &   1.0170    &   2.8175  &   3.5044  \\
7    &   0.9679    &   2.4252  &   3.9411  \\
8    &   0.9334    &   2.0929  &   4.9716  \\
9    &   0.9282    &   2.0330  &   5.2793  \\  
10   &   0.9316    &   2.0732  &   5.0596  \\
11   &   0.9320    &   2.0781  &   5.0330  \\  \bottomrule
\end{tabular} 
}
\label{Table 3}
\end{table}

\newpage

\begin{center}
{\large{\bf Table 4}}
\end{center}

\begin{table}
\tbl{Virial-expansion coefficients $c_n$ for power-law potentials with the powers
$p=6$ and $p=12$.}    
{\begin{tabular}{lll} \toprule
$c_n$~~~~~~  & $p=6$~~~~~~  & $p=12$   \\ \midrule
$c_1$    &   3.712219       &  2.566507   \\
$c_2$    &   5.551999       &  3.79107      \\
$c_3$    &   1.44261        &  3.52761     \\
$c_4$    &  -1.68834        &  2.1149    \\
$c_5$    &   1.8935         &  0.7695  \\ 
$c_6$    &  -1.700          &  0.0908  \\
$c_7$    &   0.44           & -0.074   \\ \bottomrule
\end{tabular} }
\label{Table 4} 
\end{table}

\newpage

\begin{center}
{\large{\bf Table 5}}
\end{center}

\begin{table}
\tbl{Self-similar factor approximants for the compressibility 
factor $Z_k^*(y)$ of soft-sphere fluid with $p=6$. In the bottom,
Monte Carlo results are given.}    
{\begin{tabular}{lllll} \toprule
$y$~~~~~~  &   $0.1  $~~~~~~  & $0.8$~~~~~~ & $1.6$~~~~~~~ & $2.2$   \\ \midrule
$Z_2^*$    &   1.4311         & 10.4301     & 51.9228  & 133.2476  \\
$Z_3^*$    &   1.4290         &  9.2900     & 37.8526  &  83.7437 \\
$Z_4^*$    &   1.4280         &  7.8688     & 22.8531  &  39.4063 \\
$Z_5^*$    &   1.4280         &  7.9311     & 23.6476  &  41.7210 \\
$Z_6^*$    &   1.4280         &  7.9106     & 23.3287  &  40.7276\\ 
$Z_7^*$    &   1.4280         &  7.9052     & 23.2200  &  40.3585 \\ \midrule
$Z_{MC}$   &   1.4280         &  7.902      & 23.236   &  40.43 \\ \bottomrule
\end{tabular} }
\label{Table 5} 
\end{table}

\newpage

\begin{center}
{\large{\bf Table 6}}
\end{center}

\begin{table}
\tbl{Self-similar factor approximants $Z_k^*(y)$ for the compressibility factor of 
soft-sphere fluid with $p=12$. In the bottom, Monte Carlo results are given.}    
{\begin{tabular}{lllll} \toprule
$y$~~~~~~  &   $0.1  $~~~~~~  & $0.5$~~~~~~ & $0.9$~~~~~~~ & $1.1$   \\ \midrule
$Z_2^*$    &   1.2992         &  4.1641     & 17.1316  &  39.6640  \\
$Z_3^*$    &   1.2984         &  3.8856     & 11.8805  &  20.6706 \\
$Z_4^*$    &   1.2983         &  3.8246     & 10.6982  &  17.1346 \\
$Z_5^*$    &   1.2983         &  3.8292     & 10.8221  &  17.5262 \\
$Z_6^*$    &   1.2983         &  3.8295     & 10.8324  &  17.5623 \\ 
$Z_7^*$    &   1.2983         &  3.8293     & 10.8259  &  17.5384 \\ \midrule
$Z_{MC}$   &   1.2983         &  3.825      & 10.802   &  17.456 \\ \bottomrule
\end{tabular} }
\label{Table 6} 
\end{table}

\newpage

\begin{figure}[ht]
\centerline{
\includegraphics[width=7cm]{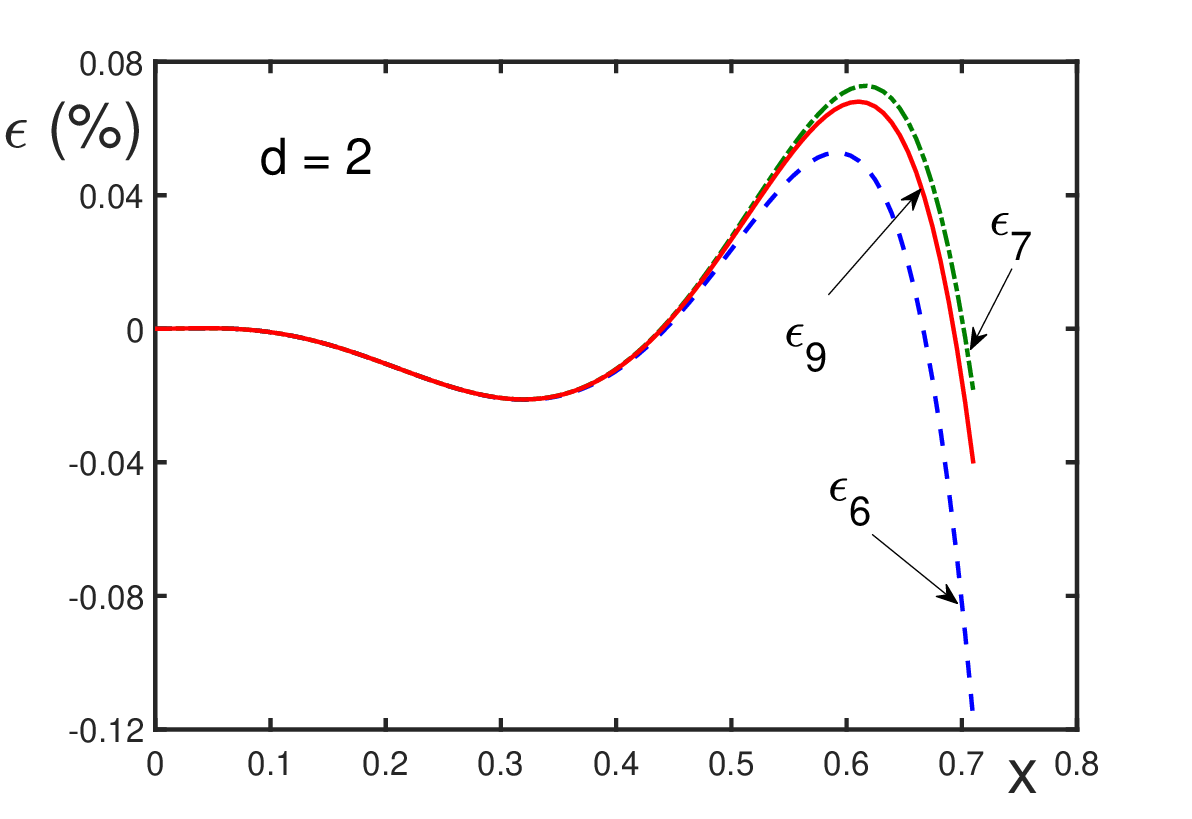}  }
\caption{\small
Relative difference, in percent, between the self-similar
approximations for compressibility factors $Z_k^*(x)$ and the equation of
state (\ref{23}) for the stable hard-disk fluid, as a function of the
packing fraction $x$.}
\label{fig:Fig.1}
\end{figure}

\vskip 2cm

\begin{figure}[ht]
\centerline{
\hbox{
\includegraphics[width=6.5cm]{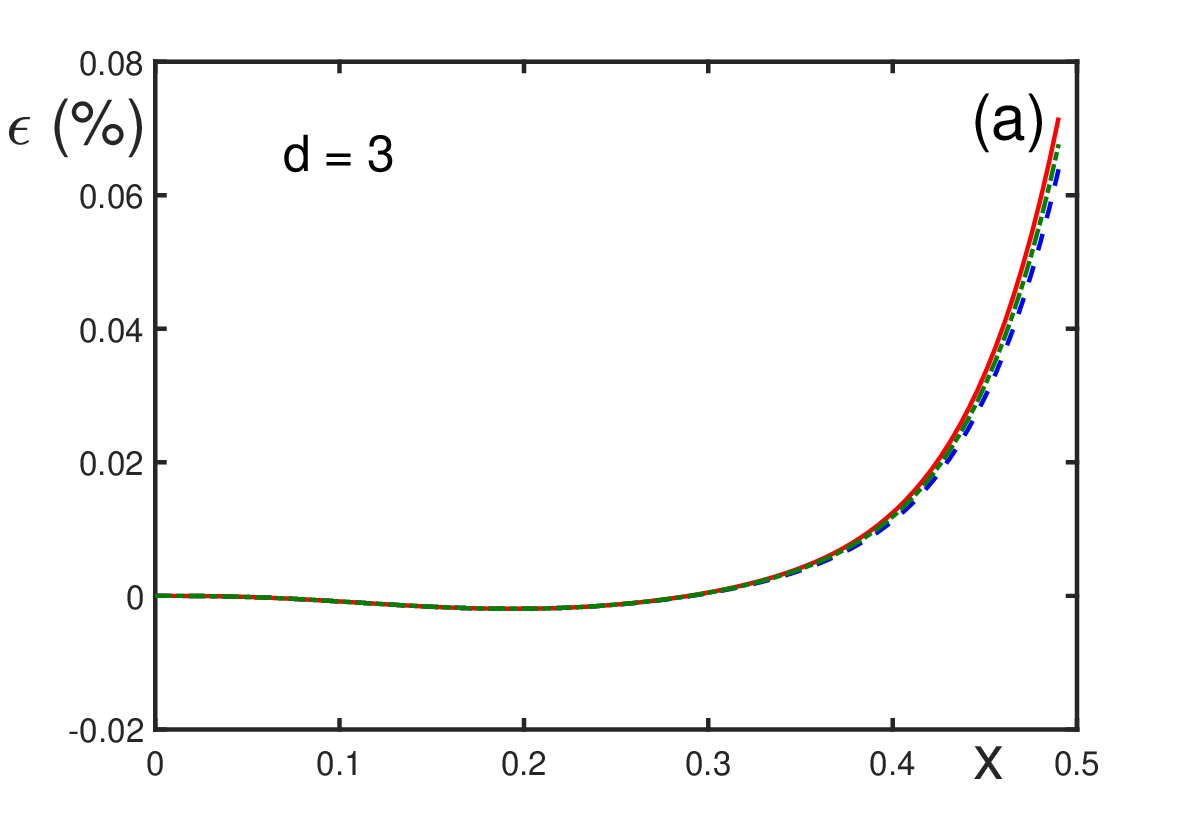} \hspace{1cm}
\includegraphics[width=6.5cm]{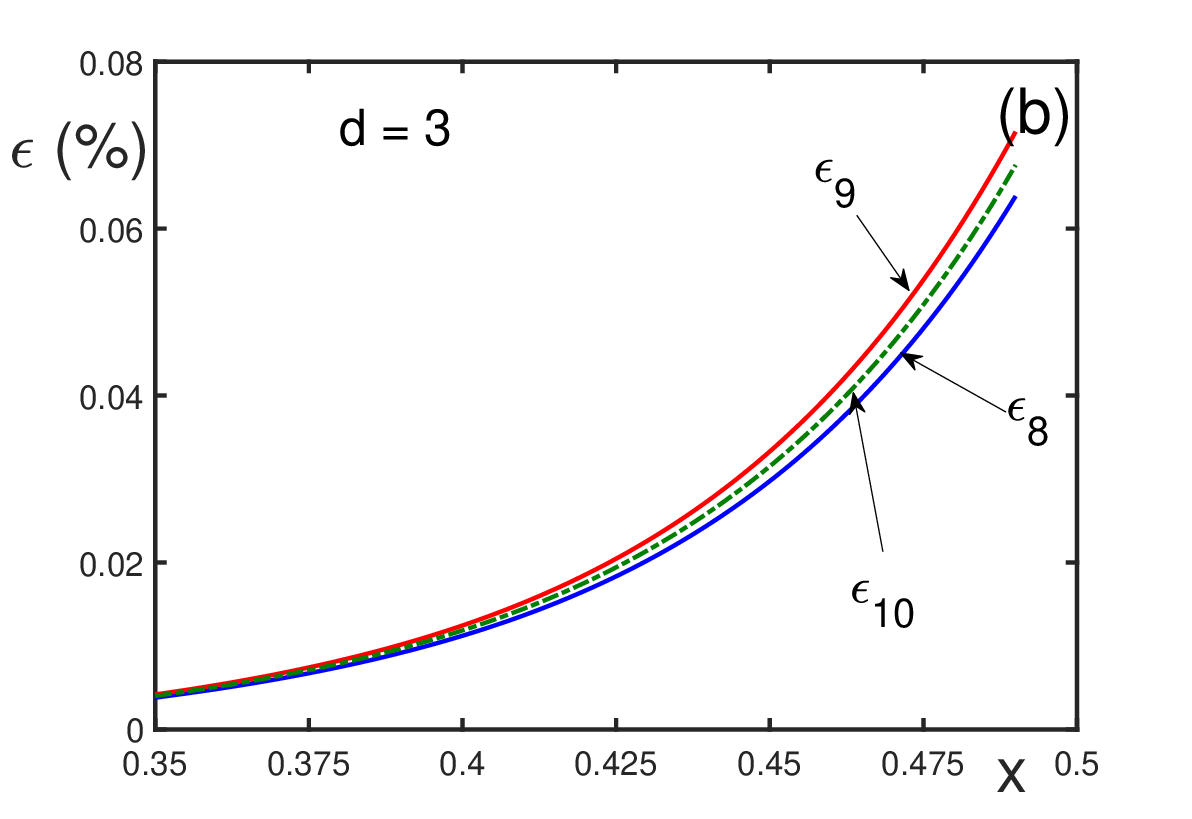} } }
\caption{\small
Relative difference, in percent, between the self-similar approximants for
the compressibility factors $Z_k^*(x)$ and the equation of states (\ref{28}) for
the stable hard-sphere fluid, as a function of the packing fraction $x$:
(a) the whole stable region;
(b) the region before freezing. }
\label{fig:Fig.2}
\end{figure}

\newpage

\begin{center}
{\Large{\bf Figure Captions} }
\end{center}

\vskip 0.5cm
{\bf Figure 1}: Relative difference, in percent, between the self-similar
approximations for compressibility factors $Z_k^*(x)$ and the equation of
state (\ref{23}) for the stable hard-disk fluid, as a function of the
packing fraction $x$.

\vskip 1cm
{\bf Figure 2}: Relative difference, in percent, between the self-similar 
approximants for the compressibility factors $Z_k^*(x)$ and the equation 
of states (\ref{28}) for the stable hard-sphere fluid, as a function of 
the packing fraction $x$:
(a) the whole stable region;
(b) the region before freezing.

\end{document}